\documentclass[prd,aps,nofootinbib,showpacs,10pt]{revtex4}
\usepackage{amsmath,graphicx,color,epsfig,amssymb}

\begin{document} 
\title{ \hfill{\small DESY 11-201}\\[0.5em]
Analysis of $B\to K^*_J (\to  K \pi) \mu^+\mu^-$ in the higher kaon resonance region}
\author{ Cai-Dian L\"u $^a$ and Wei Wang $^b$~\footnote{Email: wei.wang@desy.de}}

\affiliation{
 \it  $^a$ Institute of High Energy Physics, P.O. Box 918(4), Beijing 100049, People's Republic of China\\
 \it  $^b$ Deutsches Elektronen-Synchrotron DESY, Hamburg 22607, Germany}

\begin{abstract}
We study the resonant contributions in the process of $\overline
B^0\to  K^- \pi^+ \mu^+\mu^-$ with the $K^-\pi^+$ invariant mass
square $m_{K\pi}^2\in [1, 5] {\rm GeV}^2$.
 Width effects of the involved strange mesons,
 $K^*(1410), K_0^*(1430), K_2^*(1430), K^*(1680), K_3^*(1780)$ and $K_4^*(2045)$, are incorporated.   In terms of helicity amplitudes, we derive a compact form for the full angular distributions, through which the branching ratios, forward-backward asymmetries and polarizations are attained.  We propose that the uncertainties in the $B\to K^*_J$ form factors can be pinned down by the measurements of a set of SU(3)-related processes.  Using results from the large energy limit, we derive the dependence of branching fractions on the $m_{K\pi}$, and  find that the $K^*_2$ resonance has a clear signature,  in particular, in the transverse polarizations.
\end{abstract}
\maketitle

\section{Introduction}\label{section:introduction}

It is anticipated that the LHC is able to answer some of the
fundamental questions in particle physics. One of great interests
is in dertermining whether the new degrees of freedom are relevant for the phenomena at the
TeV scale.  On the  one hand, many new particles have 
signatures different from the standard model (SM) particles, and measurements
of their production and decays at the LHC may provide definitive evidence on their existence. On the other hand,   low energy
processes may also be influenced by them. Rare $B$ decays, with tiny
decay probabilities in the SM, are highly  sensitive to the new
degrees of freedom and thus can be exploited as indirect searches
of these unknown effects. In particular,  $b\to sl^+l^-$ especially
$B\to K^*(\to K\pi)l^+l^-$
provide a wealth of information on the weak interactions,  in terms of a number of observables ranging
from the decay probabilities, forward-backward asymmetries (FBAs),
polarizations to a full angular analysis. The small branching
fraction, of  order $10^{-6}$ for $B\to K^*l^+l^-$, is
compensated by the high luminosity  at the $B$
factories and hadron
colliders~\cite{Aubert:2008ps,:2009zv,Aaltonen:2008xf}.  It is anticipated that the
measurements by the  LHCb detector will allow to probe the
short-distance physics at an unprecedented level and will provide 
good sensitivity to discriminate between the SM and different models
of new physics.   For instance, results by the LHCb   based on the
data with the integrated luminosity $0.3
fb^{-1}$~\cite{LHC-B-Kstarll} are in good agreement with the theory
predictions~\cite{Bobeth:2011gi}, which has placed a stringent
constraint on new physics (NP) models.

In our previous work~\cite{Li:2010ra}, we have explored the $B\to K_2^* l^+l^-$ decay mode in the
SM and two specific NP scenarios using the $B\to K_2^*$ form factors 
calculated in Ref.~\cite{Wang:2010ni}.  We provided a comprehensive analysis of the branching ratio,
FBAs, transversity amplitudes, and full angular distributions. It is pointed out that the
 $B\to K_2^*l^+l^-$ decay has several advantages in different aspects and is complementary to
the  commonly-studied mode $B\to K^*l^+l^-$. The  process $B\to K_2^*l^+l^-$ has also
  received
  considerable attention  in the SM and several variants of it in
  Refs.~\cite{Rai Choudhury:2006gv,Choudhury:2009fz,Hatanaka:2009sj,Katirci:2011mt,Junaid:2011bh,Aliev:2011gc}. On the experimental side, however, its usefulness is challenged by the ``pollution" from several other strange resonances in this mass region namely $K^*(1410), K^*_0(1430),  K^*(1680), K_3^*(1780)$ and $K_4^*(2045)$~\cite{Amsler:2008zz}. The different contributions can be separated by a partial wave analysis when a large amount of data is available, but it is necessary to explore the interference effects in physical quantities, such as the branching ratios to have a benchmark  for the possible measurement in the first running of LHC.  The aim of the present work is to achieve this goal. To do so, we will study  the $B\to K_J^* l^+l^-\to K\pi l^+l^-$ with the invariant mass square  $m_{K\pi}$ ranging from $1$ to 5 $\rm {GeV}^2$.  Using the helicity amplitudes technique we will derive a compact form for the angular distributions and a number of other quantities.
To reduce the uncertainties in $B\to K^*_J$ transition form factors,
we will propose to measure a set of useful but SU(3)-related
channels. In terms of the results derived from the large energy
symmetry, we will show the differential distributions and their
dependence on $m_{K\pi}^2$. We  further point out that for
$m_{K\pi}^2\simeq 2 {\rm GeV}^2$,  the $K^*_2$ dominates especially in the transverse polarization while at
$m_{K\pi}^2\simeq 3 {\rm GeV}^2$, the $B\to K^*(1680)$ contribution
is the largest.

The rest of the paper is organized as follows. In
Sec.~\ref{section:elements}, we give the theoretical framework
including the effective Hamiltonian and the hadronic form factors.
Sec.~\ref{sec:differentialdecaydistribution} is devoted to the derivation of the differential decay
distributions and the integrated quantities. Sec.~\ref{sec:results} is devoted to the numerical predictions in the SM. We conclude in the last section. The appendix contains our derivation of the angular distributions.

\section{Theoretical Framework}\label{section:elements}
%

The decay amplitude for $B\to K_J^*(\to K\pi)l^+l^-$ consists of two
separate parts: the short distance physics and the long-distance
physics. The former arises from the degrees of freedom higher than
$m_b$, and thus can be computed by perturbation
theory.  The low-energy effect is
usually parameterized in terms of heavy-to-light form factors.


The $b\to sl^+l^-$ effective Hamiltonian
 \begin{eqnarray}
 {\cal
 H}_{\rm{eff}}=-\frac{G_F}{\sqrt{2}}V_{tb}V^*_{ts}\sum_{i=1}^{10}C_i(\mu)O_i(\mu)\nonumber\label{eq:Hamiltonian}
 \end{eqnarray}
involves the  four-quark and the magnetic penguin operators $O_i$ and their
explicit forms can be found in Ref.~\cite{Buchalla:1995vs}.   Here $C_i(\mu)$ are the Wilson coefficients for
these local operators $O_i$, and we use the leading logarithmic values~\cite{Buchalla:1995vs}, which are listed in Table \ref{tab:wilsons}, $G_F$
is the Fermi constant,
$V_{tb}=0.999176$ and $V_{ts}=-0.03972$~\cite{Amsler:2008zz} are the
CKM matrix elements. The double Cabibbo suppressed terms,
proportional to $V_{ub}V_{us}^*$, have been omitted. $m_b=4.67^{+0.18}_{-0.06}$GeV
and $m_s=0.101^{+0.029}_{-0.021}$GeV are the $b$ and $s$ quark masses~\cite{Amsler:2008zz}.

 \begin{table}
 \caption{Wilson coefficients $C_i(m_b)$ in the leading
logarithmic approximation, with $m_W=80.4\mbox{GeV}$, $\mu=m_{b,\rm
pole}$~\cite{Buchalla:1995vs}.}
 \label{tab:wilsons}
 \begin{center}
 \begin{tabular}{c c c c c c c c c}
 \hline\hline
 \ \ \ $C_1$ &$C_2$ &$C_3$ &$C_4$ &$C_5$ &$C_6$ &$C_7^{\rm{eff}}$ &$C_9$ &$C_{10}$       \\
 \ \ \ $1.107$   &$-0.248$   &$-0.011$    &$-0.026$    &$-0.007$    &$-0.031$    &$-0.313$    &$4.344$    &$-4.669$    \\
 \hline\hline
 \end{tabular}
 \end{center}
 \end{table}

With the neglect of QCD corrections, only the operators $O_{7\gamma}, O_{9}$ and $O_{10}$ contribute to the decay amplitudes 
\begin{eqnarray}
i {\cal M}(b\to
 sl^+l^-)&=&\frac{iG_F}{\sqrt2}\frac{\alpha_{\rm em}}{\pi}V_{tb}V_{ts}^*\times
 \left( \frac{C_9+C_{10}}{4}[\bar sb]_{V-A}[\bar ll]_{V+A}
 +\frac{C_9-C_{10}}{4}[\bar sb]_{V-A}[\bar ll]_{V-A}\right. \nonumber\\
 &&\left.+ C_{7L}m_b[\bar s i\sigma_{\mu\nu}
 (1+\gamma_5)b]\frac{q^\mu}{q^2}\times[\bar l \gamma^\nu l]+ C_{7R}m_b[\bar s i\sigma_{\mu\nu}
 (1-\gamma_5)b]\frac{q^\mu}{q^2}\times[\bar l \gamma^\nu
 l]\right),\label{eq:decay-amplitude-bsll-LR}
\end{eqnarray}
where $C_{7L}=C_7$ and $C_{7R}=\frac{m_s}{m_b}C_{7L}$.  On the other hand, the operators $O_1-O_6$ also contribute starting from the one loop diagrams. The
factorizable loop terms  can be incorporated into
the Wilson coefficients $C_7$ and $C_{9}$, and thus it is convenient to
define the Wilson coefficients combinations $C_7^{\rm{eff}}$ and
$C_9^{\rm{eff}}$~\cite{Buras:1994dj}
\begin{eqnarray}
 C_7^{\rm{eff}}&=&C_7-C_5/3-C_6,\nonumber\\
 C_9^{\rm{eff}}(q^2)&=&C_9(\mu)+h(\hat{m_c},\hat{s})C_0-\frac{1}{2}h(1,\hat{s})(4C_3+4
 C_4+3C_5+C_6)\nonumber\\
 &&-\frac{1}{2}h(0,\hat{s})(C_3+3
 C_4) + \frac{2}{9}(3C_3 + C_4 +3C_5+ C_6),\label{eq:C7C9eff}
\end{eqnarray}
with $\hat{s}=q^2/m_b^2$, $C_0=C_1+3C_2+3C_3+C_4+3C_5 +C_6$, and
$\hat{m}_c=m_c/m_b$.  The auxiliary functions used  above are
\begin{eqnarray}
 h(z,\hat s)&=& -\frac{8}{9}\ln \frac{m_b}{\mu}-\frac{8}{9}\ln
 z+\frac{8}{27}+\frac{4}{9}x-\frac{2}{9}(2+x)|1-x|^{1/2}
 \left\{\begin{array}{c}
 \ln\left| \frac{\sqrt{1-x}+1}{\sqrt{1-x}-1}\right|-i\pi \;\;\; {\rm for} \;\; x\equiv \frac{4z^2}{\hat s}<1 \\
  2{\rm arctan}\frac{1}{\sqrt{x-1}}\;\;\; {\rm for} \;\; x\equiv \frac{4z^2}{\hat s}>1
 \end{array}\right.,\nonumber\\
 h(0,\hat s)&=& -\frac{8}{9}\ln \frac{m_b}{\mu}-\frac{4}{9}\ln
 \hat s+\frac{8}{27}+\frac{4}{9}i\pi.
\end{eqnarray}
In the following, we shall also drop the superscripts for $C_{9}^{\rm
eff}$ and $C_{7}^{\rm eff}$ for brevity.


\begin{table}
\caption{Properties of the resonances $K_J^{*}$. The isospin symmetry relation ${\cal B} (K_J^{*} \to K^-\pi^+)=2/3{\cal B} (K_J^{*} \to K\pi)$ will be used.  }
 \label{Tab:properties}
 \begin{center}
 \begin{tabular}{ c c c ccc c c}
\hline \hline
 $K_J^{*}$       & $J^P$   & $n^{2S+1}L_J$  & $m$ (MeV) & $\Gamma$ (MeV) &${\cal B} (K_J^{*} \to K\pi)(\%)$                & $\alpha_L$ & $\beta_T$     \\
 \hline
\hline
 $K^*(1410)$    & $1^-$ & $2^3S_1?$ & $1414\pm 15$ & $232\pm 21$ & $ 6.6\pm 1.3$ & $1$ & 1 \\ \hline
 $K^*_0(1430)$    & $0^+$ & $1^3P_0,2^3P_0?$ & $1425\pm 50$ & $270\pm 80$ & $ 93\pm 10$
& 1 &--  \\ \hline
 $K^*_2(1430)$    & $2^+$ & $^3P_2$ & $1432.4\pm 1.3$ & $109\pm 5$ & $ 49.9\pm 1.2$
& $\sqrt{ \frac{2}{3}}$& $\sqrt{ \frac{1}{2}}$ \\ \hline
 $K^*(1680)$    & $1^-$ & $1^3D_1$ & $1717\pm 27$ & $322\pm 110$ & $ 38.7\pm 2.5$ & 1& 1\\
 \hline
 $K^*_3(1780)$    & $3^-$ & $1^3D_3$ & $1776\pm 7$ & $159\pm 21$ & $18.8\pm 1.0$& $\sqrt{ \frac{2}{5}}$& $\sqrt{ \frac{4}{15}}$\\ \hline
 $K^*_4(2045)$    & $4^+$ & $1^3F_4$ & $2045\pm 9$ & $198\pm 30$ & $9.9\pm 1.2$ & $\sqrt{ \frac{8}{35}}$& $\sqrt{ \frac{1}{7}}$\\ \hline
 \end{tabular}
 \end{center}
 \end{table}

\begin{table}
\caption{ $B\to K_J^{*}$ form factors taken from Ref.~\cite{Hatanaka:2009sj}. }
 \label{Tab:formfactors}
 \begin{center}
 \begin{tabular}{ c c c ccc c c}
\hline \hline
 $K_J^{*}$       & $\xi_{||}$  & $\xi_{\perp}$    \\
 \hline
\hline
 $K^*(1410)$    & $ 0.22\pm 0.03$ & $0.28\pm 0.04$ \\ \hline
 $K^*_0(1430)$    & $ 0.22\pm 0.03$ & --  \\ \hline
 $K^*_2(1430)$    & $0.22\pm 0.03$ &$0.28\pm 0.04$  \\ \hline
 $K^*(1680)$    & $ 0.18\pm 0.03$ & $0.24\pm 0.05$ \\
 \hline
 $K^*_3(1780)$    & $0.16\pm 0.03$ & $0.23\pm 0.05$ \\ \hline
 $K^*_4(2045)$    & $0.13\pm 0.03$ & $0.19\pm 0.05$ \\ \hline
 \end{tabular}
 \end{center}
 \end{table}

 The $B\to K_0^*(1430)$ transition form factors are defined by
  \begin{eqnarray}
    \langle K^*_0(P_2) |\bar s  \gamma_\mu\gamma_5  b|\overline  {B}(P_{B})\rangle
    &=&-i\left\{\left
    [P_\mu - \frac{m_{B}^2-m_{K^*_0}^2}{q^2}q_\mu \right ]
    F_1 (q^2)
    +\frac{m_{B}^2-m_{K^*_0}^2}{q^2}q_\mu F_0 (q^2)\right\} ,\nonumber\\
   \langle  K^*_0(P_2) |\bar s  \sigma_{\mu\nu} q^\nu\gamma_5  b|\overline  {B}(P_{B})\rangle &=&\left[ (m_{B}^2-m_{K^*_0}^2) q_\mu - q^2 P_\mu\right ]
  \frac{F_T (q^2)}{
   m_{B}+m_{K^*_0}},
   \end{eqnarray}
while the parametrization of the $B\to K_J^*(J \geqslant 1)$ form factors is as follows~\cite{Hatanaka:2009sj,Yang:2010qd} 
 \begin{eqnarray}
  \langle K_J^*(P_2,\epsilon)|\bar s\gamma^{\mu}b|\overline B(P_B)\rangle
   &=&-\frac{2V(q^2)}{m_B+m_{K_J^*}}\epsilon^{\mu\nu\rho\sigma} \epsilon^*_{J\nu}  P_{B\rho}P_{2\sigma}, \nonumber\\
  \langle  K_J^*(P_2,\epsilon)|\bar s\gamma^{\mu}\gamma_5 b|\overline
  B(P_B)\rangle
   &=&2im_{K_J^*} A_0(q^2)\frac{\epsilon^*_{J } \cdot  q }{ q^2}q^{\mu}
    +i(m_B+m_{K_J^*})A_1(q^2)\left[ \epsilon^*_{J\mu }
    -\frac{\epsilon^*_{J } \cdot  q }{q^2}q^{\mu} \right] \nonumber\\
    &&-iA_2(q^2)\frac{\epsilon^*_{J} \cdot  q }{  m_B+m_{K_J^*} }
     \left[ P^{\mu}-\frac{m_B^2-m_{K_J^*}^2}{q^2}q^{\mu} \right],\nonumber\\
  \langle  K_J^*(P_2,\epsilon)|\bar s\sigma^{\mu\nu}q_{\nu}b|\overline
  B(P_B)\rangle
   &=&-2iT_1(q^2)\epsilon^{\mu\nu\rho\sigma} \epsilon^*_{J\nu} P_{B\rho}P_{2\sigma}, \nonumber\\
  \langle  K_J^*(P_2,\epsilon)|\bar s\sigma^{\mu\nu}\gamma_5q_{\nu}b|\overline  B(P_B)\rangle
   &=&T_2(q^2)\left[(m_B^2-m_{K_J^*}^2) \epsilon^*_{J\mu }
       - {\epsilon^*_{J } \cdot  q }  P^{\mu} \right] +T_3(q^2) {\epsilon^*_{J } \cdot  q }\left[
       q^{\mu}-\frac{q^2}{m_B^2-m_{K_J^*}^2}P^{\mu}\right],\nonumber\label{eq:BtoTformfactors-definition}
 \end{eqnarray}
which is in general analogous to
the $B\to K^*$
ones. Here $q=P_B-P_2$, and $P=P_B+P_2$. We have the relation
$2m_{K_J^*}A_0(0)=(m_B+m_{K_J^*})A_1(0)-(m_B-m_{K_J^*})A_2(0)$ in
order to smear the pole at $q^2=0$. The polarization vector in the above equations is constructed by the $J$-rank polarization tensor
\begin{eqnarray}
  &&\epsilon_{J\mu}(h) =\frac{1}{m_B^{J-1}}
  \epsilon_{\mu\nu_1 \nu_2 ...\nu_{J-1}}(h)P_{B}^{\nu_1}P_{B}^{\nu_2}...P_{B}^{\nu_{J-1}},
\end{eqnarray}
with $h=0,\pm1$ being the helicity.
Using the expression for $\epsilon_{\mu\nu_1 \nu_2 ...\nu_{J-1}}(h)$ which is a product of the polarization vectors with the Clebsch-Gordan coefficients, we simplify the above equation as $\epsilon_{J\mu}(h)\sim (|\vec p_{K^*_J}| /m_{K^*_J})^{J-1} \tilde \epsilon_{J\mu}$, with $\tilde \epsilon_{J\mu}(0)= \alpha_L^J \epsilon_\mu(0)$  and $\tilde \epsilon_{J\mu}(\pm 1)= \beta_T^J \epsilon_\mu(\pm 1)$ and $|\vec p_{K^*_J}|\sim E_{K^*_J}$ in the large recoil region. $\alpha_L^J$ and $\beta_T^J$ are products of the Clebsch-Gordan coefficients
\begin{eqnarray}
 \alpha_L^J &=& C^{J,0}_{1,0;J-1,0} C^{J-1,0}_{1,0; J-2,0} ... C^{2,0}_{1,0;1,0},\nonumber\\
 \beta_T^J &=& C^{J,1}_{1,1;J-1,0} C^{J-1,0}_{1,0; J-2,0} ... C^{2,0}_{1,0;1,0}.
\end{eqnarray}

The $B\to K^*_J$ form factors are nonperturbative in nature and the
application of QCD theory to them  mostly resorts to the Lattice QCD
simulations, which is quite limited at this stage.  The crucial
input we use in this work is the observation that, in the heavy
quark $m_b\to \infty$ and the large energy $E\to \infty$ limit,
interactions of the heavy and light systems can be expanded in small
ratios $\Lambda_{QCD}/E$ and $\Lambda_{QCD}/m_B$. At the leading
power, the large energy symmetry is obtained and  such symmetry to a
large extent simplifies the heavy-to-light transition~\cite{HUTP-90-A071,hep-ph/9812358}. As a concrete
application,  the current $\bar s\Gamma b$ in QCD can be matched
onto the current $\bar s_n \Gamma b_v$ constructed in terms of the
fields in the effective theory. Here $v$ denotes the velocity of the
heavy meson and $n$ is a light-like vector along the $K^*_J$ moving
direction. This procedure constrains the independent Lorentz
structures and reduces the seven independent hadronic form factors
for each $B\to K^*_J$ ($J\ge1$)  type to two universal functions
$\xi_\perp$ and $\xi_{||}$. Explicitly, we
have 
\begin{eqnarray}
 A_0^{K^*_J}(q^2) \left(\frac{|\vec p_{K^*_J}|}{m_{K^*_J}}\right)^{J-1}\equiv A_0^{K^*_J, \rm eff}\simeq (1-\frac{m_{K^*_J}^2}{m_B E}) \xi_{||}^{K^*_J}(q^2) +\frac{m_{K^*_J}}{m_B} \xi_{\perp}^{K^*_J}(q^2),\nonumber\\
 A_1^{K^*_J}(q^2) \left(\frac{|\vec p_{K^*_J}|}{m_{K^*_J}}\right)^{J-1}\equiv A_1^{K^*_J, \rm eff}\simeq
 \frac{2E}{m_B+m_{K^*_J}} \xi_{\perp}^{K^*_J}(q^2),\nonumber\\
 A_2^{K^*_J}(q^2) \left(\frac{|\vec p_{K^*_J}|}{m_{K^*_J}}\right)^{J-1}\equiv A_2^{K^*_J, \rm eff}\simeq
 (1+\frac{m_{K^*_J}}{m_B})[\xi_{\perp}^{K^*_J}(q^2) -\frac{m_{K^*_J}}{E}\xi_{||}^{K^*_J}(q^2)],\nonumber\\
V^{K^*_J}(q^2) \left(\frac{|\vec p_{K^*_J}|}{m_{K^*_J}}\right)^{J-1}\equiv V^{K^*_J, \rm eff}\simeq
 (1+\frac{m_{K^*_J}}{m_B}) \xi_{\perp}^{K^*_J}(q^2),\nonumber\\
T_1^{K^*_J}(q^2) \left(\frac{|\vec p_{K^*_J}|}{m_{K^*_J}}\right)^{J-1}\equiv T_1^{K^*_J, \rm eff}\simeq\xi_{\perp}^{K^*_J}(q^2),\nonumber\\
T_2^{K^*_J}(q^2) \left(\frac{|\vec p_{K^*_J}|}{m_{K^*_J}}\right)^{J-1}\equiv T_2^{K^*_J, \rm eff}\simeq
 (1-\frac{q^2}{m_B^2-m_{K^*_J}^2}) \xi_{\perp}^{K^*_J}(q^2),\nonumber\\
 T_3^{K^*_J}(q^2) \left(\frac{|\vec p_{K^*_J}|}{m_{K^*_J}}\right)^{J-1}\equiv T_3^{K^*_J, \rm eff}\simeq
\xi_{\perp}^{K^*_J}(q^2) - (1-\frac{m_{K^*_J}^2}{m_B^2}) \frac{m_{K^*_J}}{E} \xi_{||}^{K^*_J}(q^2).\label{eq:LEETrelation}
\end{eqnarray}
For the sake of simplicity we will use the latter set of form factors but as in the case of $C_9$ and $C_7$, we drop the superscript ``eff"  as well.
In the case of $B$ to scalar meson transition, the large energy limit gives
\begin{eqnarray}
 \frac{m_{B}}{m_{B}+m_{K^*_0}} F_T(q^2) = F_1(q^2) = \frac{m_B}{2E} F_0(q^2)=\xi^{K^*_0}(q^2).
\end{eqnarray}
The results for $\xi_{||}^{K^*_J}$ and $\xi_{\perp}^{K^*_J}$ derived
from the Bauer-Stech-Wirbel (BSW) model~\cite{Wirbel:1985ji} in
Ref.~\cite{Hatanaka:2009sj} will be used in this work and we collect these results in Tab.~\ref{Tab:formfactors}. For the $B\to
K^*_0$ transition,  it is plausible to employ $\xi^{B\to K_0^*}= \xi_{||}^{B\to
K_2^*}$ since both $K^*_0$ and $K^*_2$ are p-wave states.

\begin{table}
\caption{$B\to K^*_2$ form factors at  $q^2=0$ in the ISGW2 model~\cite{Scora:1995ty} (using the updated inputs~\cite{Cheng:2010yd}), the
covariant light-front quark model~\cite{Cheng:2010yd,Cheng:2009ms}
and the light-cone QCD sum rules~\cite{Yang:2010qd} and perturbative QCD approach~\cite{Wang:2010ni}.}
\begin{center}
\begin{tabular}{cccccccc}
\hline \hline 
 \hline &
 ISGW2~\cite{Cheng:2010yd}     & CLFQM~\cite{Cheng:2010yd,Cheng:2009ms}   & LCSR~\cite{Yang:2010qd} & LEET+BSW~\cite{Hatanaka:2009sj} & PQCD~\cite{Wang:2010ni} \\\hline  
  $V^{BK_2^*}$ &  0.38  & 0.29 & $0.16\pm 0.02$ & $0.21 \pm 0.03$ & $0.21^{+0.06}_{-0.05}$ \\  
      $A_0^{BK_2^*}$                  &$0.27$ & 0.23 & $0.25\pm 0.04$ & $0.15\pm 0.02$ &$0.18^{+0.05}_{-0.04}$                 \\
       $A_1^{BK_2^*}$                 & 0.24 & 0.22 & $0.14\pm 0.02$ & $0.14\pm 0.02$ & $0.13^{+0.04}_{-0.03}$            \\
      $A_2^{BK_2^*}$              &$0.22$ & 0.21 & $0.05\pm 0.02$ &$0.14\pm 0.02$ & $0.08^{+0.03}_{-0.02}$               \\
        $T_1^{BK_2^*} $     &    
             &$0.28$               & $0.14\pm 0.02$  & $0.16\pm 0.02$     &$   0.17_{  -0.04         }^{+   0.05        }
 $      \\
        $T_3^{BK_2^*}$                & &$-0.25$   & $0.01^{+0.02}_{-0.01}$ & $0.10\pm 0.02$  &$   0.14
            _{  -0.03         }
            ^{+   0.05        }
 $            \\ 
\hline \end{tabular}\label{Tab:BtoK2formfactorcomparision}
\end{center}
\end{table}

Several remarks on the form factors are given in order. 

\begin{itemize}

\item Due to the lack of Lattice QCD simulations, the calculation of $B\to K^*_J$ form factors rely on different phenomenological  models.   In Tab.~\ref{Tab:BtoK2formfactorcomparision},  as an example we show the results for the $B\to K^*_2$ form factors at  $q^2=0$ in the ISGW2 model~\cite{Scora:1995ty} (using the updated inputs~\cite{Cheng:2010yd}), the
covariant light-front quark model~\cite{Cheng:2010yd,Cheng:2009ms}
and the light-cone QCD sum rules~\cite{Yang:2010qd} and perturbative QCD approach~\cite{Wang:2010ni} (using the light meson's light-cone distribution
amplitudes~\cite{Cheng:2010hn}). From this table we can see the LEET+BSW results used here are close to the ones in the light-cone sum rules  (except  for $T_3$) and the perturbative QCD approach.

\item The large energy effective theory~\cite{HUTP-90-A071,hep-ph/9812358} has neglected the interaction between the soft sector and collinear sector and it is  refined by the soft-collinear effective theory~\cite{Bauer:2000ew,Bauer:2000yr,Bauer:2001cu,hep-ph/0109045,hep-ph/0211069}, in which at leading power in $1/m_b$ a form factor takes the generic expression
\begin{eqnarray}
 F_i (q^2)= C_i \xi(q^2) +C_i' \int d\tau \Xi_{a} (\tau, q^2). 
\end{eqnarray}
Here $C_i$ and $C_i'$ are the short-distance Wilson coefficients obtained by integrating out degrees of freedom with virtuality of ${\cal O}(m_b^2)$.  $\xi$ is one of the above universal functions entering the large-recoil symmetries. $\Xi_a(\tau, q^2)$ is a symmetry breaking function, which can be factorized into a convolution of light-cone distribution amplitudes with the jet function. The detailed expressions of the $B\to K^*_J$ form factors (with the subscript ``eff") have a similar form with the $B\to V$ transition, for instance,  as Eqs.(21, 22) in Ref.~\cite{Beneke:2005gs} and in particular, two relations for form factors remain
\begin{eqnarray}
 \frac{m_{B}}{m_{B}+m_{K^*_J}} V^{K^*_J} =  \frac{m_{B}+m_{K^*_J}}{2E} A_1^{K^*_J},\;\;\; T_1^{K^*_J} = \frac{m_B}{2E} T_2^{K^*_J}. \label{eq:SCETrelation}
\end{eqnarray} 
Again, the function $\xi$ can only be calculated in some nonperturbative QCD methods. The calculation in light-cone sum rules in conjunction with the soft-collinear effective theory indicates that the $\xi$ dominates  in $B\to \pi$ transition while the $\Xi_a$ gives   corrections at the order of $5\%-10\%$~\cite{DeFazio:2005dx} \footnote{A direct fit of the hadronic $B\to \pi\pi$ decay data results in a large $\Xi_a^{B\to \pi}$~\cite{Bauer:2004tj}, while a numerically small $\Xi_a^{B\to \rho}$ seems to be favored by the $B\to \rho\rho$ data~\cite{Jain:2007dy}.}.  One may expect a similar size for $\Xi_{a}$ in $B\to K^*_J$ transition which will be one of the main sources of uncertainties.

\item It is noteworthy to point out that
there are ambiguities in the internal structures of $K^*_0$, thus
large discrepancies on form factors can be found in the literature.
For instance, using two different assignments of $K^*_0$, namely
p-wave states without or with one unit of radial excitation, we have
calculated the B to scalar meson form factors in perturbative QCD approach~\cite{Keum:2000ph} and  the results can
differ up to a factor of 3~\cite{Li:2008tk}.  We propose that the SU(3)-symmetry related  processes can be used to pin down the uncertainties. 
Channels of this type include  the semi-leptonic $B\to a_0(1450)l\bar \nu$, $B_s\to K^*_0l\bar \nu$ decays and the exclusive
channels $\bar B^0\to a_0^+(1450) D_s^-/D^-$, $B^-\to a_0^0(1450) D_s^-$,
$\bar B_s\to K^{*+}_0 D^-$. Semileptonic decays provide the information of form factors in  the full kinematics region through the differential decay width distribution. The above exclusive processes
are color-favored and free of annihilation diagrams;   therefore the
factorization method works phenomenologically  well for them. In the factorization context, the
decay amplitudes, taking $\bar B^0\to D_s^-a_0^+(1450)$ as an
example, are written as
\begin{eqnarray}
  {\cal A}(\bar B^0\to D_s^- a_0^+)= \frac{iG_F}{\sqrt 2}i V_{cs}^* V_{ub} a_1  f_{D_s} (m_{B}^2-m_{D_s}^2) F_0^{B\to a_0}(m_{D_s}^2),
\end{eqnarray}
where $a_1\sim 1$ being the Wilson coefficients and $f_{D_s}$
denoting the decay constant of the $D_s$ meson.  In particular, most of the
inputs will be canceled if the ratio
\begin{eqnarray}
r=\frac{\Gamma(\bar B^0\to D_s^- a_0^+)}{\Gamma(\bar B^0\to D_s^- \pi^+)}\simeq \frac{[F_0^{B\to a_0}(m_{D_s}^2)]^2}{[F_0^{B\to \pi}(m_{D_s}^2)]^2}
\nonumber
\end{eqnarray}
is considered. The decay $\bar B^0\to D_s^- \pi^+$ has  a quite large branching ratio ${\cal B}= (2.4\pm
0.4)\times 10^{-5}$~\cite{Amsler:2008zz}.  The measurement of $\bar
B^0\to D_s^- a_0^+$ in the future will consequently determine the
$F_0^{B\to a_0}$ and also   $F_0^{B\to K^*_0}$ up to
SU(3) symmetry breaking effects. Replacing $D_s^-$ by $D_s^{*-}$,
one can extract the form factor $F_1^{B\to a_0/K^*_0}$ from the relevant data in future.

\end{itemize}

\section{Differential decay distributions and forward-backward asymmetries}\label{sec:differentialdecaydistribution}

\begin{figure}\begin{center}
\includegraphics[scale=0.3]{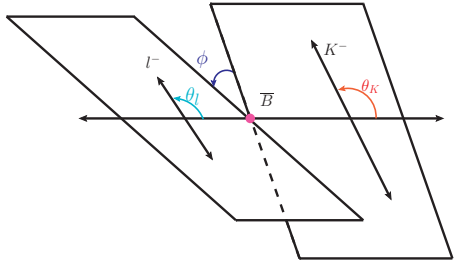}
\caption{Kinematics in  $\overline B\to  \overline K_J^*(\to
K^-\pi^+)l^+l^-$. $K_J^*$ moves along the $z$ axis in the $B$ rest frame.  $\theta_K(\theta_l)$ is defined in $K^*_J$ (lepton pair) rest frame as the angle between $z$-axis and the flight direction of $K^-$ ($\mu^-$), respectively.   The azimuth angle $\phi$ is the angle between the $K^*_J$ decay and lepton pair planes. } \label{fig:angles}
\end{center}
\end{figure}

The convention on the kinematics in $B\to K_J^*(\to K\pi)l^+l^-$ is illustrated in Fig.~\ref{fig:angles}. The moving
direction of $K_J^*$ in the $B$ rest frame is chosen as the $z$
axis. The polar angle $\theta_K$ ($\theta_l$) is defined as the
angle between the flight direction of $K^-$ ($\mu^-$) and the $z$
axis in the $K_J^*$ (lepton pair) rest frame. $\phi$ is the angle
defined by the decay planes of $K_J^*$ and the lepton pair.

$B\to K_J^*(\to K\pi)l^+l^-$ is a quasi four-body decay process and proceeds
via three steps: $B$ meson first decays into a nearly onshell strange
meson plus a pair of leptons; the $K_J^*$ meson  propagates followed
by its strong decay into the $K\pi$ state.
The decay amplitudes of $B\to (K^-\pi^+)l^+l^-$ are obtained by
sandwiching
Eq.~\eqref{eq:decay-amplitude-bsll-LR} between the initial and final
states, in which the spinor product $[\bar sb]$ by hadronic
matrix element will be replaced by hadronic form factors.
 The operator realization of this picture is
\begin{eqnarray}
 \langle l^+l^-| [\bar l l]|0\rangle  \langle K\pi | [\bar s b] |\overline B^0\rangle &\simeq & \langle l^+l^-| [\bar l l]|0\rangle  \int d^4 p_{K^*_J} \frac{ \langle K\pi|K^*_J\rangle \langle K^*_J | [\bar s b] |\overline B^0\rangle }{ p_{K^*_J}^2- m_{K^*_J}^2 +i m_{K^*_J} \Gamma_{K^*_J}},
\end{eqnarray}
with $p^2_{K^*_J}=m^2_{K\pi}$.
In appendix~A, we will compute the required quantities in the three steps with the use  of  helicity amplitudes.
Combining the individual pieces, we obtain the angular distributions
\begin{eqnarray}
 \frac{d^4\Gamma}{dm_{K\pi}^2dq^2d\cos\theta_K d\cos\theta_l d\phi}&=& \Big[I_1^c  + 2I_1^s +(I_2^c +2I_2^s ) \cos(2\theta_l) + 2I_3 \sin^2\theta_l
 \cos(2\phi)+2\sqrt 2I_4\sin(2\theta_l)\cos\phi \nonumber\\
 && +2\sqrt 2I_5 \sin(\theta_l) \cos\phi+2I_6 \cos\theta_l+2\sqrt 2I_7 \sin(\theta_l) \sin\phi\nonumber\\
 && +
 2\sqrt 2I_8\sin(2\theta_l)\sin\phi+2I_9 \sin^2\theta_l
 \sin(2\phi)\Big],
\end{eqnarray}
with the angular coefficients
\begin{eqnarray}
 I_1^c&=&  (|A_{L0}|^2+|A_{R0}|^2)
 +8 \frac{m_l^2}{q^2}{\rm Re}[A_{L0}A^*_{R0} ]+4\frac{m_l^2}{q^2} |A_t|^2, \nonumber\\
 I_1^s&=&\frac{3}{4} [|A_{L\perp}|^2+|A_{L||}|^2+|A_{R\perp}|^2+|A_{R||}|^2  ]
 \left(1-\frac{4m_l^2}{3q^2}\right)+\frac{4m_l^2}{q^2} {\rm Re}[A_{L\perp}A_{R\perp}^*
 + A_{L||}A_{R||}^*],\nonumber\\
 I_2^c  &=& -\beta_l^2(  |A_{L0}|^2+ |A_{R0}|^2),\nonumber\\
 I_2^s  &=&
 \frac{1}{4}\beta_l^2(|A_{L\perp}|^2+|A_{L||}|^2+|A_{R\perp}|^2+|A_{R||}|^2),
 \nonumber\\
 I_3  &=&\frac{1}{2}\beta_l^2(|A_{L\perp}|^2-|A_{L||}|^2+|A_{R\perp}|^2-|A_{R||}|^2),\nonumber\\
 I_4
  &=& \frac{1}{\sqrt2}\beta_l^2
  [{\rm Re}(A_{L0}A_{L||}^*)+{\rm
  Re}(A_{R0}A_{R||}^*],\;\;\;\;\;\;\;\;
 I_5
  = \sqrt 2\beta_l
  [{\rm Re}(A_{L0}A_{L\perp}^*)-{\rm Re}(A_{R0}A_{R\perp}^*)],\nonumber\\
 I_6  &=& 2\beta_l
  [{\rm Re}(A_{L||}A^*_{L\perp})-{\rm
  Re}(A_{R||}A^*_{R\perp})],\;\;\;\;\;
 I_7
 = \sqrt2\beta_l
  [{\rm Im}(A_{L0}A^*_{L||})-{\rm Im}(A_{R0}A^*_{R||})],\nonumber\\
 I_8 &=& \frac{1}{\sqrt2}\beta_l^2
  [{\rm Im}(A_{L0}A^*_{L\perp})+{\rm
  Im}(A_{R0}A^*_{R\perp})],\;\;\;\;\;
 I_9
 =\beta_l^2
  [{\rm Im}(A_{L||}A^*_{L\perp})+{\rm
  Im}(A_{R||}A^*_{R\perp})].\label{eq:angularCoefficients}
\end{eqnarray}
The lepton mass correction factor is $\beta_l=\sqrt{1-4m_l^2/q^2}$.  The functions $A_{L/Ri}$ are defined by
\begin{eqnarray}
 A_{L/R 0/t }&=& \sum_{J=0,1,2...}  \sqrt{ N_{K_J^*}}Y_{J}^0(\theta,0){\cal M}_B(K^*_J, L/R, 0/t ) \frac{i}{ m_{K\pi}^2- m_{K^*_J}^2+i m_{K^*_J} \Gamma_{K^*_J}}    \sqrt\frac{m_{K^*_J} \Gamma_{K^*_J\to K\pi} }{ \pi},\nonumber\\
 A_{L/R ||/\perp }&=& \sum_{J=0,1,2...}  \sqrt{ N_{K_J^*}}Y_{J}^{-1}(\theta,0){\cal M}_B(K^*_J, L/R, ||/\perp) \frac{i}{ m_{K\pi}^2- m_{K^*_J}^2+i m_{K^*_J} \Gamma_{K^*_J}}     \sqrt\frac{m_{K^*_J} \Gamma_{K^*_J\to K\pi} }{ \pi},\nonumber
\end{eqnarray}
with $ N_{K_J^*} =\frac{\sqrt {\lambda} {q^2}\beta_l}{256\pi^3 m_B^3}$. ${\cal M}_B$ is the decay amplitudes of $B\to K^*_J V$ to be given in the appendix. 
In the narrow-width limit,  the integration over the $K\pi$ invariant mass will be conducted as
\begin{eqnarray}
 \int dm_{K\pi}^2 \frac{m_{K^*_J}\Gamma_{K^*_J}}{\pi} \frac{1}{ (m_{K\pi}^2-m_{K^*_J}^2)^2+m_{K^*_J}^2\Gamma_{K^*_J}^2}=1.
\end{eqnarray}


Integrating out the angles $\theta_l,\theta_K$ and $\phi$, we obtain
the dilepton mass spectrum
\begin{eqnarray}
 \frac{ d^2\Gamma_L}{dq^2 dm_{K\pi}^2}
 &=& \frac{2}{3} \int_{-1}^1 2\pi d\cos\theta_K  \left(3I_1^c- I_2^c\right),\nonumber\\
 \frac{ d^2\Gamma_T}{dq^2 dm_{K\pi}^2}
 &=& \frac{2}{3} \int_{-1}^1 2\pi d\cos\theta_K  \left(6I_1^s-2I_2^s\right),\nonumber\\
 \frac{ d^2\Gamma}{dq^2 dm_{K\pi}^2}
 &=&  \frac{ d^2\Gamma_L}{dq^2 dm_{K\pi}^2}+ \frac{ d^2\Gamma_T}{dq^2 dm_{K\pi}^2},
\end{eqnarray}
and its expression in the massless limit
 \begin{eqnarray}
 \frac{ d^2\Gamma_i}{dq^2 dm_{K\pi}^2}
 &=&   \frac{8}{3}  \int_{-1}^1 2\pi d\cos\theta_K  (|A_{Li}|^2+|A_{Ri}|^2),
 \end{eqnarray}
with $i=0,\pm1$ or $i=0,\perp,||$.


The differential FBA in this process is defined by
\begin{eqnarray}
 \frac{d^2 A_{FB}}{dq^2dm_{K\pi}^2}&=&\left[\int_0^1 -\int_{-1}^0\right] d\cos\theta_l\frac{d^3\Gamma}{dq^2 d\cos\theta_l dm_{K\pi}^2}
 = \frac{2}{3} \int_{-1}^1 2\pi \cos\theta_K 3 I_6,
\end{eqnarray}
while  the normalized differential FBA is
given by
\begin{eqnarray}
 \frac{\overline {d^2A_{FB}}}{dq^2dm_{K\pi}^2}&=&\frac{\frac{d^2 A_{FB}}{dq^2dm_{K\pi}^2}}
 { \frac{d^2\Gamma}{dq^2dm_{K\pi}^2}}=
 \frac{ \int_{-1}^1 d\cos\theta_K 3I_6}{\int_{-1}^1 d\cos\theta_K [3I_1^c+6I_1^s-I_2^c-2I_2^s]}.
\end{eqnarray}

 \section{Numerical Results and Discussions}\label{sec:results}

Before presenting the numerical results,
 we start with an estimate of the contributions from different mesons. It is noticed that  larger the  $J$ is the smaller is the contribution. 
(1) The $K^*_J$ with a larger spin is heavier and thus the phase
space is smaller. (2) The Clebsch-Gordan coefficient products
$\alpha_L^J$ and $\beta_T^J$ decrease with  the increase of $J$. (3) The $B\to K^*_J$
form factors suppress the heavier $K^*_J$ further. Moreover, the
tiny branching ratios of $K^*(1410)$ and $K^*_4(2045)$ into
$K^-\pi^+$ result in very smaller effects. As a consequence we
find that the $K^*_4(2045)$ is negligibly small.

\begin{figure}
\begin{center}
\includegraphics[scale=0.35]{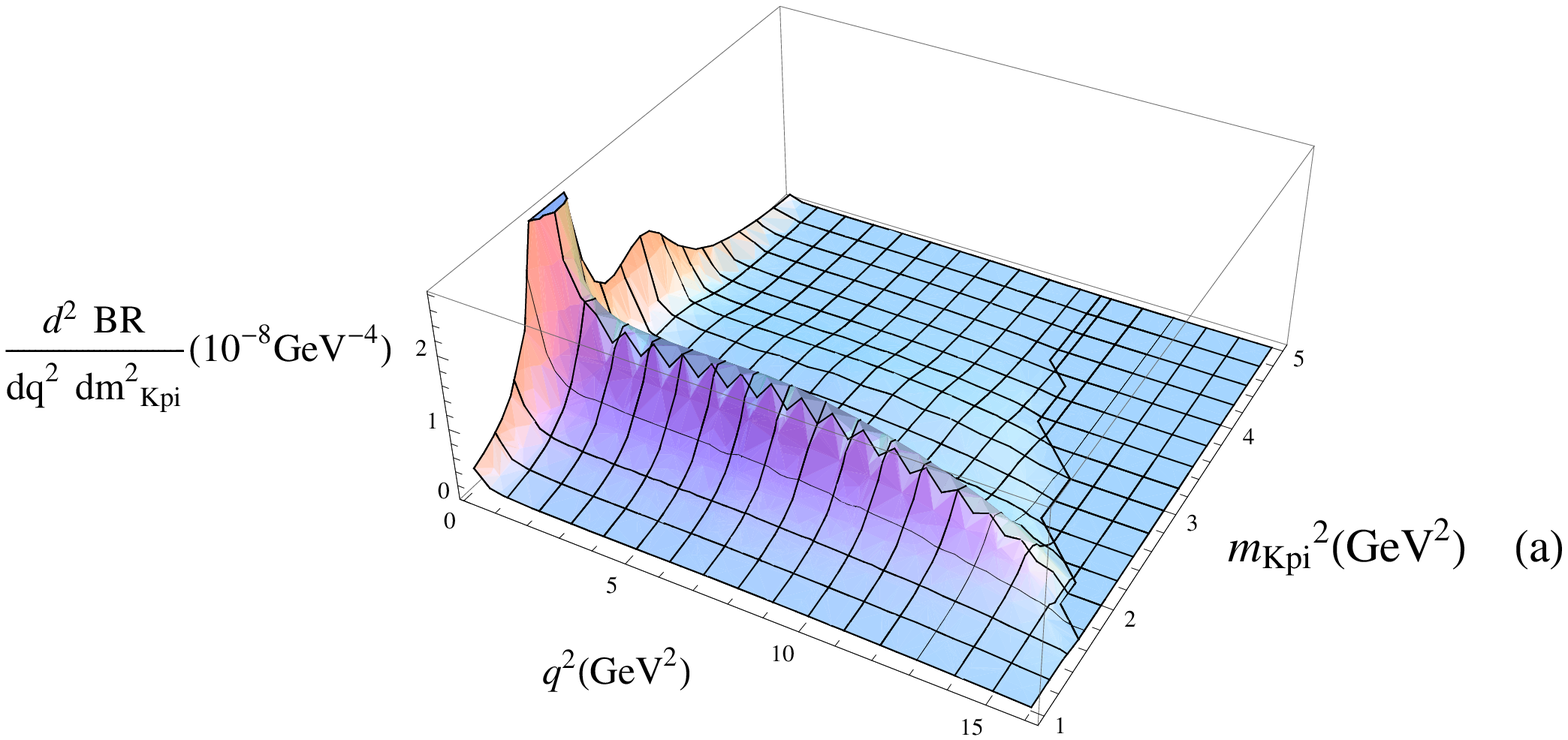}
\includegraphics[scale=0.35]{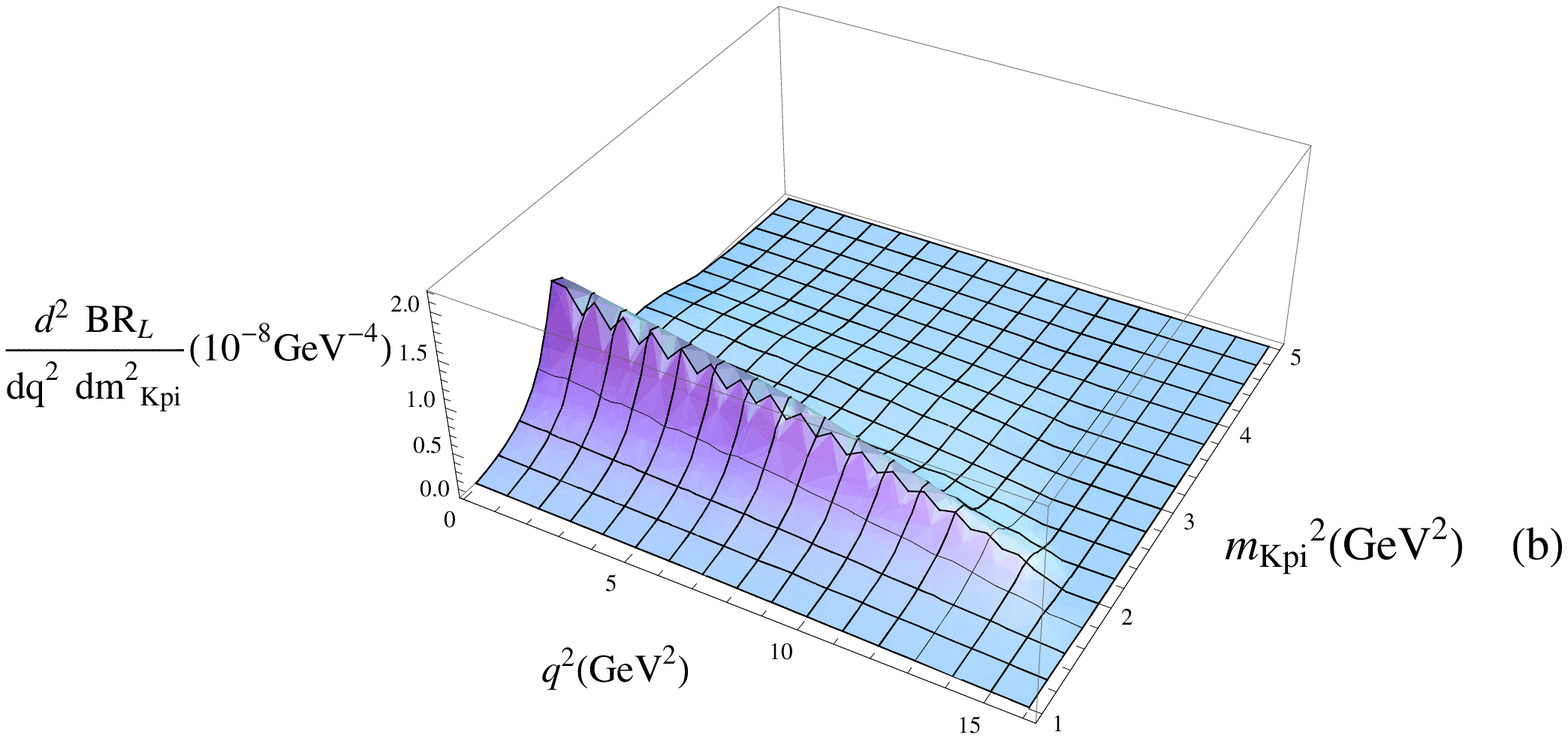}
\includegraphics[scale=0.35]{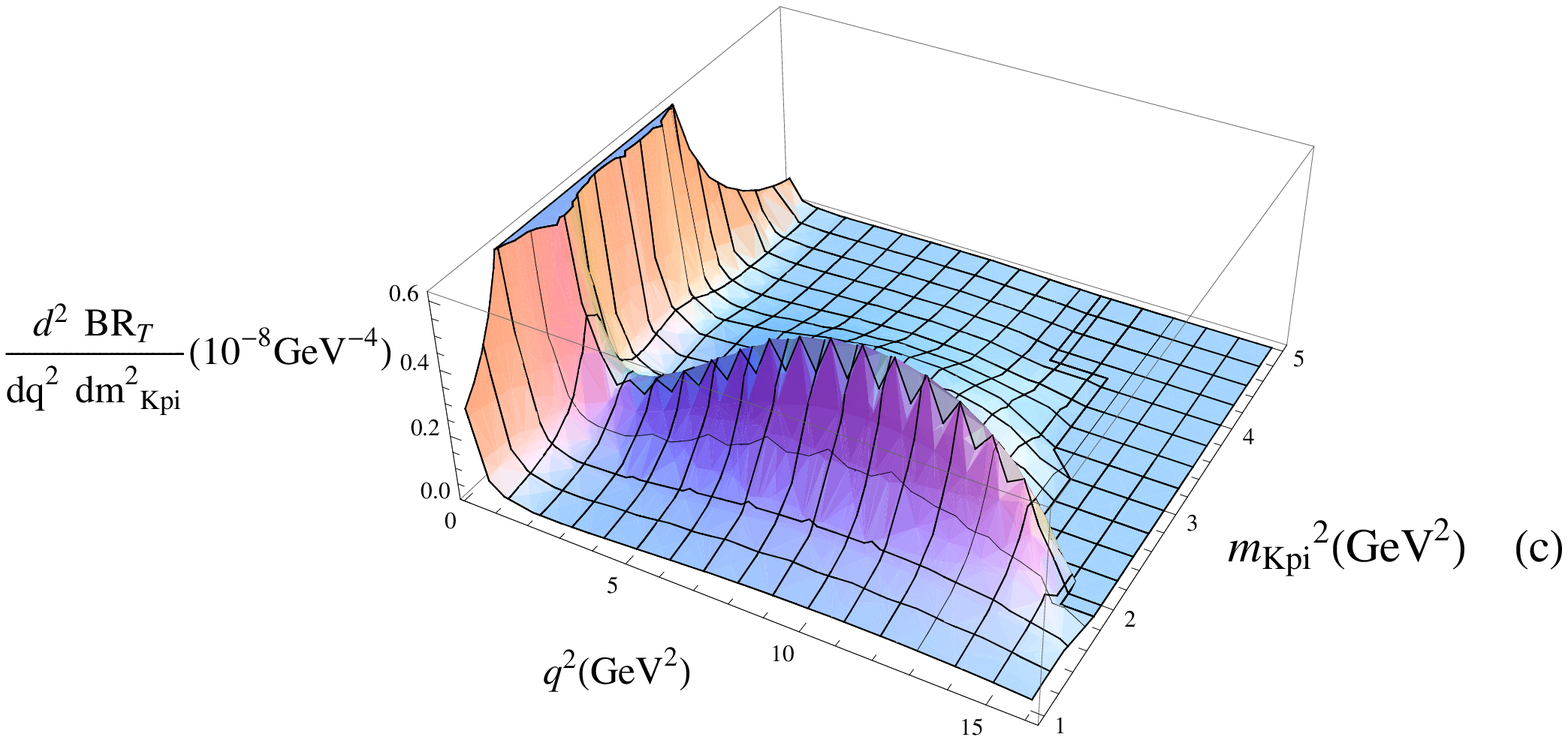}
\includegraphics[scale=0.35]{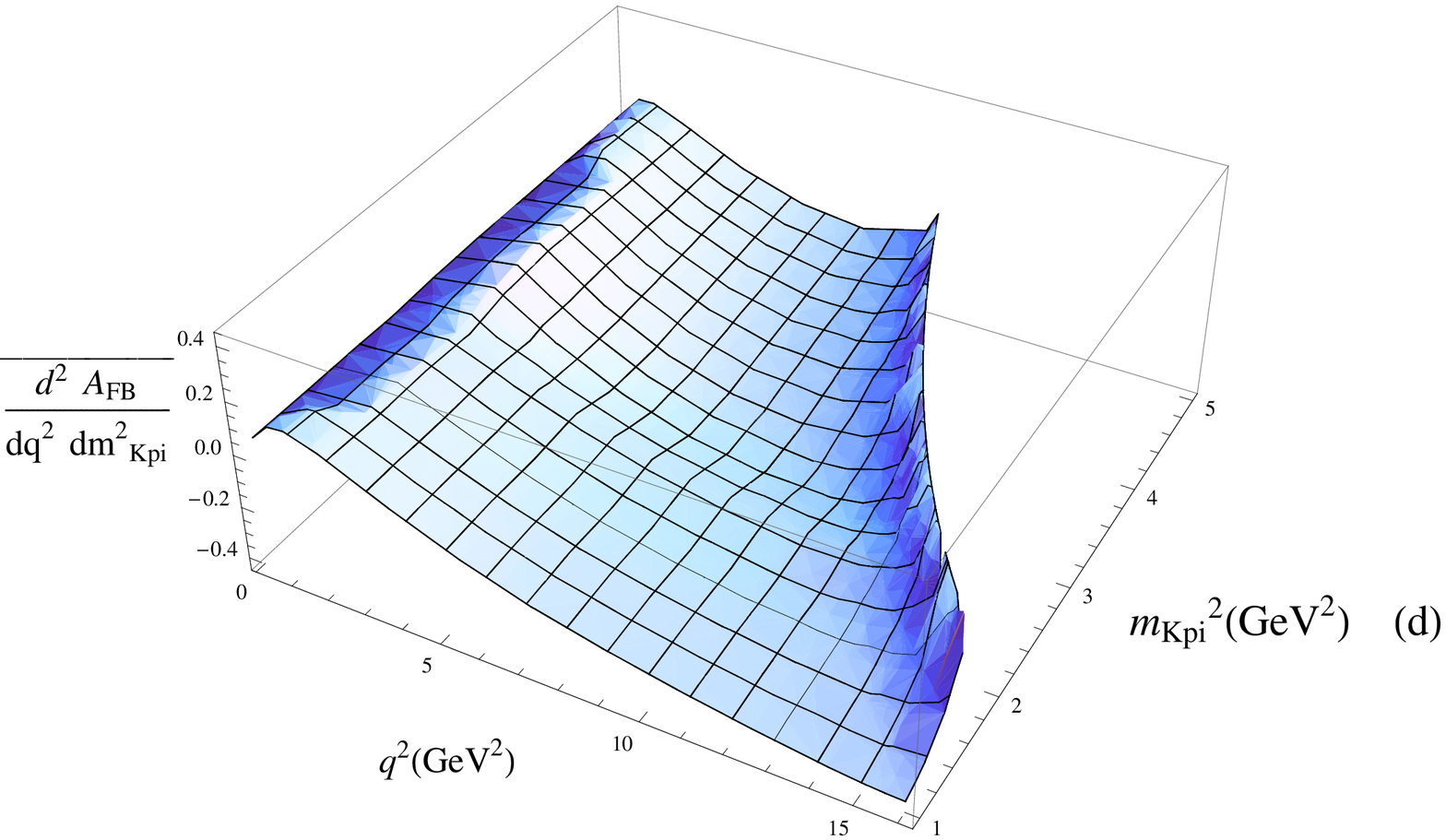}
\caption{Differential branching ratios $\frac{d^2BR_i}{ dq^2 dm_{K\pi}^2}$, with $i$ denoting the total (a), longitudinal (b) and transverse polarizations (c), and the normalized FBA $\frac{\overline {d^2A_{FB}}}{dq^2dm_{K\pi}^2}$ (d) for $\bar B^0\to K^-\pi^+ \mu^+\mu^-$ in the mass region $1 {\rm GeV}^2<m_{K\pi}^2<5 {\rm GeV}^2$}\label{fig:3Dmu}
\end{center}
\end{figure}


\begin{figure}
\begin{center}
\includegraphics[scale=0.5]{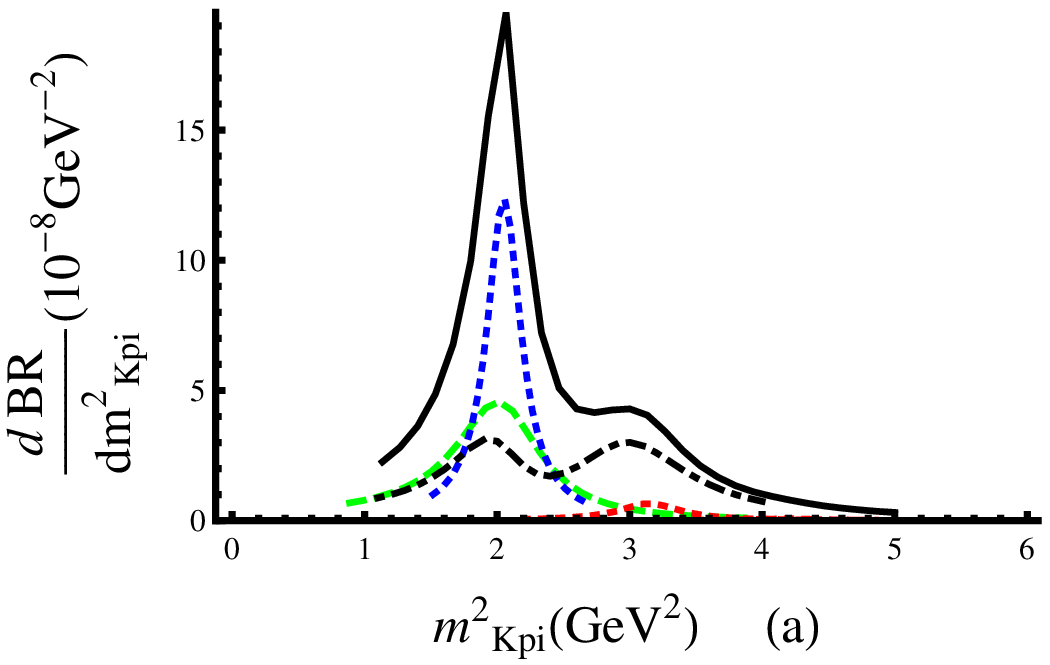}
\includegraphics[scale=0.5]{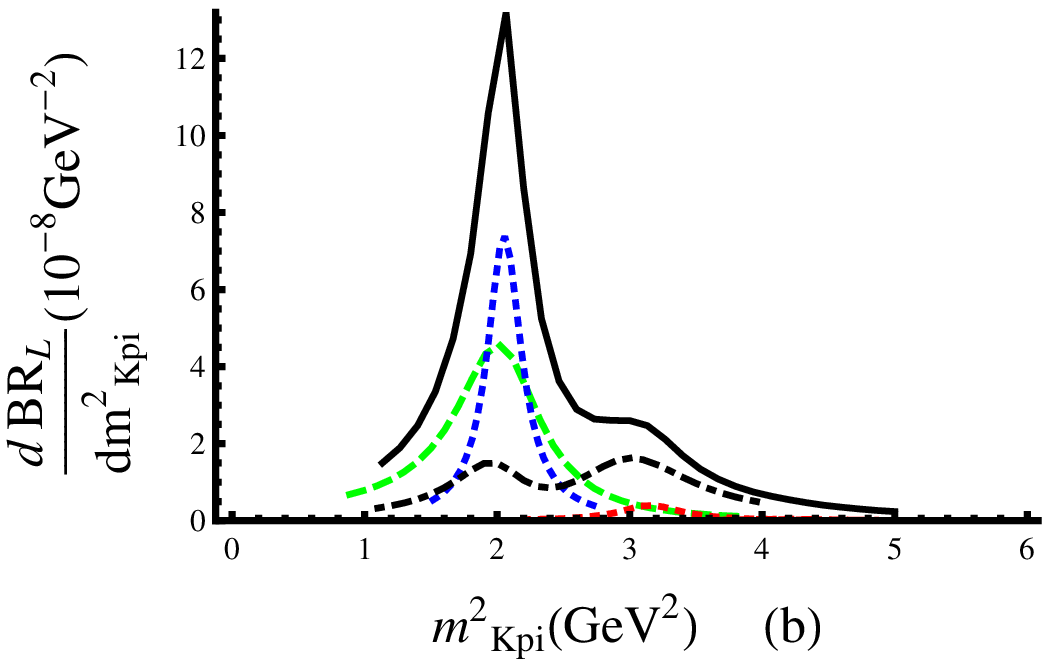}
\includegraphics[scale=0.5]{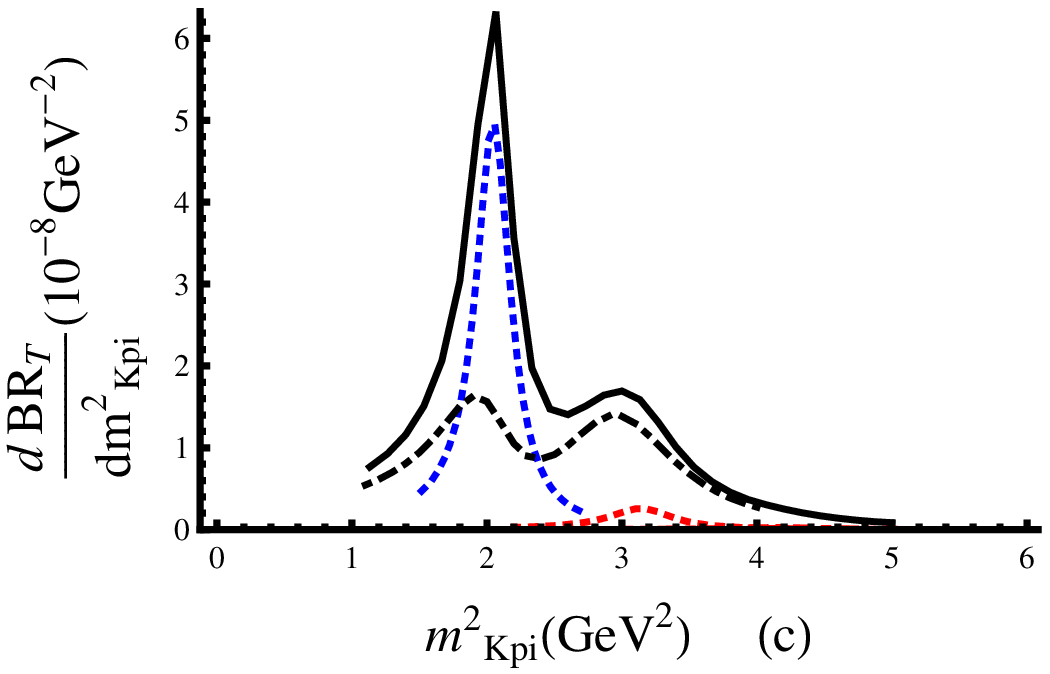}
\caption{Differential branching ratios
$\frac{dBR_i}{dm_{K\pi}^2}(\overline B^0\to K^-\pi^+\mu^+\mu^-)$
integrated over the kinematics region $q^2>4m_l^2$. The black (solid)
curve denotes the total contribution, while individual terms are
given by  the green (dashed) line  for $K^*_0(1430)$,  blue dotted line  for
$K_2^*(1430)$, the black (dot-dashed) line  for $K^*(1680)$ and $K^*(1410)$
with the  interference incorporated, and the red (dotted) curve with a very small magnitude for
$K_3^*(1780)$. The contribution from $K_4^*(2045)$ is negligibly small. }
\label{fig:all2Dmu}
\end{center}
\end{figure}

\begin{figure}
\begin{center}
\includegraphics[scale=0.5]{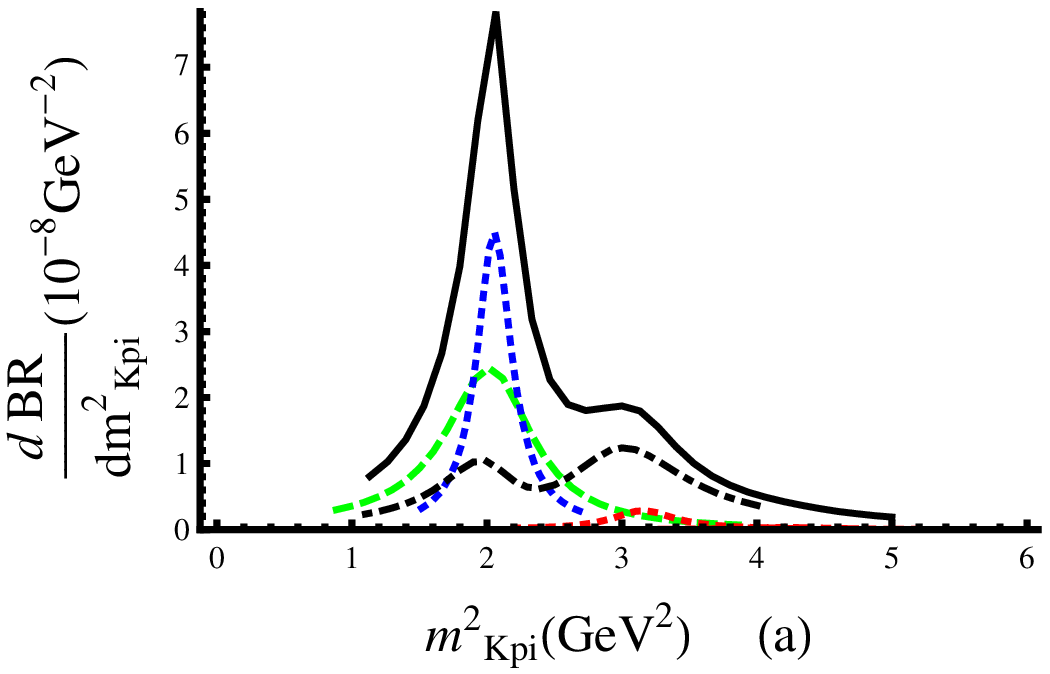}
\includegraphics[scale=0.5]{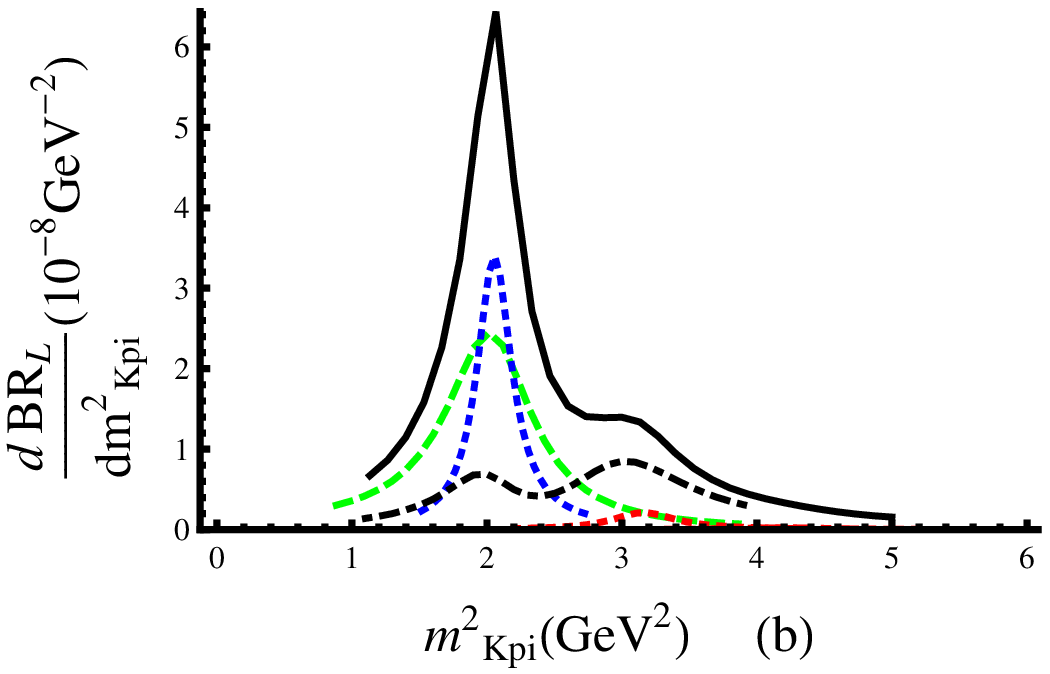}
\includegraphics[scale=0.5]{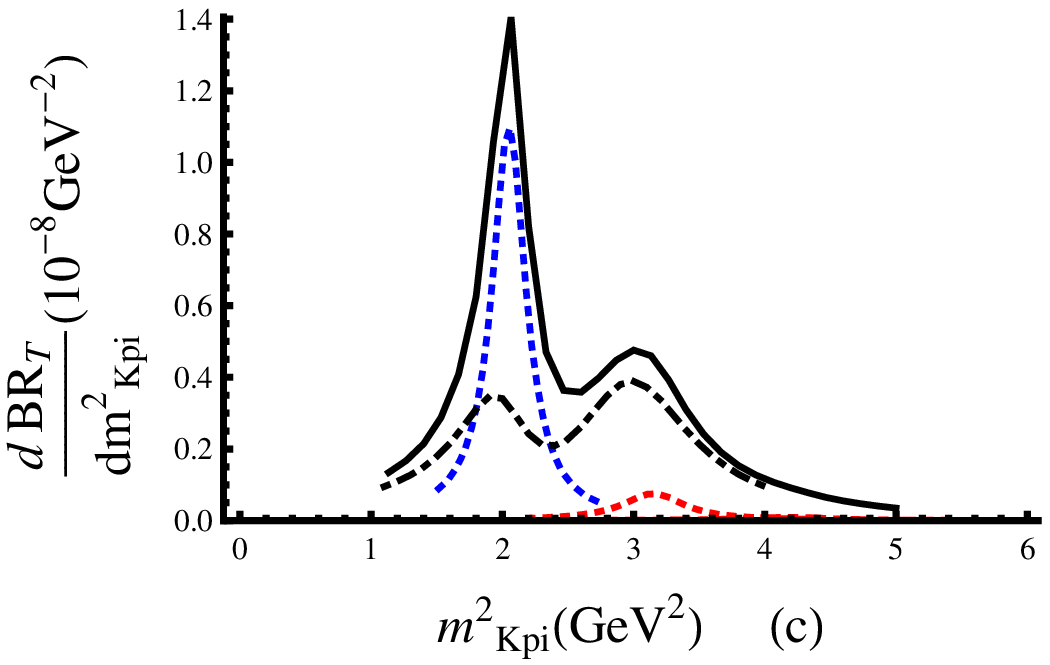}
\caption{Same as Fig.~\ref{fig:all2Dmu}, but integrated over $1{\rm GeV}^2<q^2<6{\rm GeV}^2$ }
\label{fig:162Dmu}
\end{center}
\end{figure}

We plot the differential branching ratios $\frac{d^2 BR_i}{ dq^2
dm_{K\pi}^2}$ (in units of  $10^{-8} {\rm GeV}^{-4}$),  with the
subscript $i$ denoting the total, longitudinal and transverse
polarizations) and the normalized FBA $\frac{\overline
{d^2A_{FB}}}{dq^2dm_{K\pi}^2}$ for $\bar B^0\to K^-\pi^+ \mu^+\mu^-$
in Fig.~\ref{fig:3Dmu}. By integrating the differential
distributions over $q^2$, we obtain their dependence on
$m_{K\pi}^2$. Fig.~\ref{fig:all2Dmu} shows the differential
branching ratios $\frac{dBR_i}{dm_{K\pi}^2}(\overline B^0\to
K^-\pi^+\mu^+\mu^-)$ (in units of $10^{-8} {\rm GeV}^{-2}$)
integrated over the kinematics region $4m_l^2<q^2< (m_B-
{m_{K\pi}})^2$, while Fig.~\ref{fig:162Dmu} gives the results under
the integration over $1  {\rm GeV}^2 <q^2<6  {\rm GeV}^2$. In these
figures, the black (solid)
curve denotes the total contribution, while individual terms are
given by  the green (dashed) line  for $K^*_0(1430)$,  blue dotted line  for
$K_2^*(1430)$, the black (dot-dashed) line  for $K^*(1680)$ and $K^*(1410)$
with the  interference incorporated, and the red (dotted) curve with a very small magnitude for
$K_3^*(1780)$. The contribution from $K_4^*(2045)$ is negligibly small.From these figures, we can see that for $m_{K\pi}^2=2  {\rm
GeV}^2$,  the $K^*_2$ dominates; while at  $m_{K\pi}^2\simeq 3  {\rm
GeV}^2$, the $B\to K^*(1680)$ contribution is the largest,
especially in the transverse polarization.


Now let us  analyze 
the zero crossing point $s_0$ of FBAs satisfying  $\frac{d^2 A_{FB}}{dq^2dm_{K\pi}^2}|_{q^2=s_0}=0$  and governed by   the equation 
\begin{eqnarray}
  {\rm Re} [C_9] A_1(s_0) V(s_0) +C_{7L} \frac{m_b(m_B+m_{K^*_J})}{s_0} A_1(s_0) T_1(s_0) +C_{7L} \frac{m_b(m_B-m_{K^*_J})}{s_0} T_2(s_0) V(s_0)=0.\label{eq:s0}
\end{eqnarray}
Substituting the relations from the large energy limit into the above equation, we find that the dependence on the form factors cancels completely and more explicitly Eq.~\eqref{eq:s0} gives 
\begin{eqnarray}
 s_0= (3.1\pm 0.1){\rm GeV}^2,
\end{eqnarray} 
where the uncertainties are caused by   $m_{K^*_J}^2/m_B^2$ corrections in the form factor relations in Eq.~\eqref{eq:LEETrelation}. 
As we have discussed in Sec.~II, the interaction of collinear and soft sectors brings in symmetry breaking effects. After the inclusion of them, only two relations among form factors remain as in Eq.~\eqref{eq:SCETrelation}. Define the ratio
\begin{eqnarray}
 {\cal R}^{K^*_J}(q^2)\equiv \frac{m_B+m_{K^*_J}}{m_{B}} \frac{T_1^{K^*_J}(q^2)}{ V^{K^*_J}(q^2)},
\end{eqnarray}
we find that Eq.~\eqref{eq:s0} becomes 
\begin{eqnarray}
   {\rm Re} [C_9] + 2\frac{m_b m_B}{s_0} C_{7L} {\cal R}^{K^*_J}(s_0)=0. 
\end{eqnarray}
The analaysis in Ref.~\cite{Beneke:2005gs} indicates that the ratio ${\cal R}$ can deviate from 1 by 10$\%$  in the $B\to V$ transition (see Eq.(124) of \cite{Beneke:2005gs}). Using the PQCD~\cite{Wang:2010ni}  and LCSR~\cite{Yang:2010qd} results for the $B\to K_2$ form factors, we have 
\begin{eqnarray}
 {\cal R}^{K_2}_{\rm PQCD}\sim 1.03,\;\; 
 {\cal R}^{K_2}_{\rm LCSR}\sim 1.11,
 \end{eqnarray} 
 where the $q^2$-dependence is  negligible since form factors $T_1$ and $V$ are found to have similar $q^2$-distribution in both model calculations.  
Suppose that the  ${\cal R}^{K^*_J}$ deviates from 1 by 10\%, the $s_0$ is also shifted by roughly 10\%, namely $0.3$ GeV$^2$. 

Our analysis of  the $m_{K\pi}$ dependence can be generalized to
similar channels such as $\bar B^0\to J/\psi K^*_J \to (\mu^+\mu^-)
(K^-\pi^+)$ and $B_s\to f_J (\to K^+K^-) l^+l^-$. For the former
processes, however, apart from the $B\to K^*_J$ form factors, it is
likely that the effective Wilson coefficients $a_2$ depends on the
spin of $K^*_J$ as well (for a recent discussion see
Ref.~\cite{Colangelo:2010bg}).  Although the relative strengths
among $K^*_J$ may be modified, the structure of the dependence on $m_{K\pi}^2$ is expected to be similar.  For the latter, uncertainties are presumably smaller,
as a recent measurement of $B_s\to J/\psi K^+K^-$ clearly shows the
peak at $f_2'(1525)$~\cite{LHCbBsf2}.

\section{Conclusion}

In summary,
we  have analyzed the resonant contributions in the process 
$\overline B^0\to  K^-\pi^+ \mu^+\mu^-$ with the
$K^-\pi^+$ invariant mass square $m_{K\pi}^2\in [1, 5] {\rm GeV}^2$.
Width effects of the strange mesons involved in this range,
$K^*(1410), K_0^*(1430), K_2^*(1430), K^*(1680), K_3^*(1780)$ and $
K_4^*(2045)$, are incorporated.   In terms of the helicity amplitudes,
we derive a compact form for the full angular distributions, through
which the branching ratios, forward-backward asymmetries and
polarizations are attained.  To pin down the uncertainties in the
form factors, we suggest the measurements of a set of SU(3)-related
processes which are useful.  Using the form factors from the large
energy limit, we derive the dependence of the branching fractions on 
$m_{K\pi}$,  and we point out that the  $K^*_2$ and $K^*(1680)$
contributions can be separated from the rest,  in particular, in the
transverse polarizations.  The generalization into $\bar B \to J/\psi
K^*_J(\to K^-\pi^+)$ and $B_s\to f_J(\to K^+K^-) l^+l^-$ is also discussed briefly.

\section*{Acknowledgements}

W. W. is grateful to
Ulrik Egede for the enlightening discussions on the interferences of 
$K_J^*$ which initiated this work. We  thank  Ahmed Ali   for
fruitful discussions, and Run-Hui Li for the collaboration at the early stage of this work. 
This work  is  supported by National Natural Science
Foundation of China under the Grant No. 10735080 and 11075168
  (C.D. L\"u) and the
Alexander von Humboldt Stiftung (W. Wang). 

\begin{appendix}

\section{Helicity amplitudes}

The differential distributions are divided into  several individual pieces and each of them can be expressed in terms of the helicity amplitudes which are Lorentz invariant.
\begin{itemize}
\item  $B$ decays  into $K^*_J$

The  spin-0 $K^*_0$ in the final state has only one polarization state and the longitudinal amplitudes are
\begin{eqnarray}
 i{\cal M}_B(K^*_0, L/R,0)&=& N_1 i\Bigg[ (C_9\mp C_{10}) \frac{\sqrt {\lambda}}{\sqrt{ q^2}} F_1(q^2) +2(C_{7L}-C_{7R})  \frac{\sqrt {\lambda }m_b}{\sqrt {q^2}(m_B+m_{K^*_0})}F_T(q^2) \Bigg],\nonumber\\
 i{\cal M}_B(K^*_0, L/R,t)&=&N_1 i\Bigg[ (C_9\mp C_{10}) \frac{m_B^2-m_{K^*_0}^2}{\sqrt {q^2}} F_0(q^2) \Bigg],
\end{eqnarray}
with  $N_1= \frac{iG_F}{4\sqrt 2} \frac{\alpha_{\rm em}}{\pi} V_{tb}V_{ts}^*$. The
function $\lambda$ is related to the magnitude of the $K_J^*$ momentum in
$B$ meson rest frame: $\lambda\equiv\lambda(m^2_{B},m^2_{K_J^*},
q^2)=2m_B|\vec p_{K_J^*}|$, and
$\lambda(a^2,b^2,c^2)=(a^2-b^2-c^2)^2-4b^2c^2$.  Here  the script $t$ denotes the time-like component of a virtual vector/axial-vector meson decays into a lepton pair.    In the case of strange mesons with spin $J\ge1$, the $K^-\pi^+$ system can be either longitudinally or transversely polarized:
\begin{eqnarray}
 i{\cal M}_B(K^*_J, L,0)&=&\frac{ \alpha_L^J N_1  i}{2m_{K^*_J}\sqrt {q^2}}\left[ (C_9-C_{10})  [(m_B^2-m_{K^*_J}^2-q^2)(m_B+m_{K^*_J})A_1
 -\frac{\lambda}{m_B+m_{K^*_J}}A_2]\right.\nonumber\\
 &&\left. + 2m_b(C_{7L}-C_{7R})  [ (m_B^2+3m_{K^*_J}^2-q^2)T_2 -\frac{\lambda  }{m_B^2-m_{K^*_J}^2}T_3]\right], \nonumber\\
 i{\cal M}_B({K^*_J}, L,\pm)
 &=& \beta_T^J N_1  i \left[ (C_9-C_{10}) [(m_B+m_{K^*_J})A_1\mp \frac{\sqrt \lambda}{m_B+m_{K^*_J}}V]\right.\nonumber\\
 &&\left.
 -\frac{2m_b(C_{7L}+C_{7R})}{q^2} (\pm\sqrt \lambda T_1)+\frac{2m_b(C_{7L}-C_{7R})}{q^2} (m_B^2-m_{K^*_J}^2)T_2\right],\\
 i{\cal M}_B({K^*_J}, L, t)&=&\alpha_L^J i N_{1} (C_9- C_{10})\frac{\sqrt \lambda}{\sqrt {q^2}}A_0.
\end{eqnarray}
For the sake of convenience, we define
\begin{eqnarray}
 i{\cal M}_{B}(K^*,L,\perp/||)&=&\frac{1}{\sqrt 2}[i{\cal M}_B(K^*, L,+) \mp i{\cal M}_B(K^*, L,-)],\nonumber\\
i{\cal M}_B(K^*, L,\perp) &=& -i\beta_T^J \sqrt{2} N_1\left[(C_9-C_{10})
 \frac{\sqrt \lambda V}{m_B+m_{K^*_J}}+\frac{2m_b(C_{7L}+C_{7R})}{q^2}\sqrt \lambda T_1\right],\nonumber\\
i{\cal M}_B(K^*, L,||)&=& i\beta_T^J\sqrt{2} N_{1} \left[(C_9-C_{10}) (m_B+m_{K^*_J})A_1+\frac{2m_b(C_{7L}-C_{7R})}{q^2}(m_B^2-m_{K^*_J}^2)
 T_2 \right].
\end{eqnarray}
The right-handed decay amplitudes are defined in a similar way
\begin{eqnarray}
 A_{Ri}
  &=& A_{Li}|_{C_{10}\to -C_{10}}.
\end{eqnarray}
The combination of the time-like decay amplitude is used in the
differential distribution
\begin{eqnarray}
 i{\cal M}_B(K^*_0, t)&=&i{\cal M}_B(K^*_0, R,t)-i{\cal M}_B(K^*_0, L,t)=2\alpha_L^J i C_{10}N_1   \frac{m_B^2-m_{K^*_0}^2}{\sqrt {q^2}} F_0(q^2) ,\\
 i{\cal M}_B(K^*_J, t)&=&i{\cal M}_B(K^*_J, R,t)-i{\cal M}_B(K^*_J, L,t)= 2\alpha_L^J iN_{1}  C_{10}\frac{\sqrt \lambda}{\sqrt {q^2}}A_0(q^2).
\end{eqnarray}

\item
Nonzero leptonic amplitudes are given as follows
\begin{eqnarray}
 {\cal M}_{L,R}(\lambda_l,\lambda_{\bar l},\lambda_V)=T^{L,R}_{\lambda_l,\lambda_{\bar l}} D^{1*}_{\lambda_V, \lambda_l-\lambda_{\bar l}}(\phi,\pi-\theta_l,0),\nonumber\\
{\cal M}_L(\frac{1}{2},\frac{1}{2},t)=-{\cal M}_L(\frac{-1}{2},\frac{-1}{2},t)=-
{\cal M}_R(\frac{1}{2},\frac{1}{2},t)={\cal M}_R(\frac{-1}{2},\frac{-1}{2},t)=-2m_l.
\end{eqnarray}
with $q_\pm=\sqrt {q^2}\pm\sqrt {q^2-4m_l^2}$.
The reduced matrix elements are given as
\begin{eqnarray}
T^L_{\frac{1}{2}\frac{1}{2}}=
T^L_{\frac{-1}{2}\frac{-1}{2}}=
T^R_{\frac{1}{2}\frac{1}{2}}=
T^R_{\frac{-1}{2}\frac{-1}{2}}= -2m_l ,\nonumber\\
 T^L_{\frac{1}{2} \frac{-1}{2}}= T^R_{\frac{-1}{2} \frac{1}{2}}= \sqrt 2{q_-} ,\;\;
 T^L_{\frac{-1}{2} \frac{1}{2}}=  T^R_{\frac{1}{2} \frac{-1}{2}}=-\sqrt 2 {q_+}.
\end{eqnarray}

\item The  propagation of   $K^*_J$ is parameterized  by a Breit-Wigner formula while the $K_J^*\to K\pi$ decay  is described by the spherical harmonic functions: $Y_{J}^{i}(\theta_K,0)$, with $i=0$ for $K_0^*$ and $i=0,\pm1$ for $K^*_J$.  It should be pointed out that the dependence of the coupling between a virtual  $K^*_J$ and the $K\pi$ state on $m_{K\pi}^2$ is neglected.  Since there is no singularity in the coupling, it can be expanded in terms of $m_{K\pi}^2-m_{K^*_J}^2$ around the resonance region, which is also guaranteed by the Breit-Wigner propagation. For definiteness, we list the explicit forms of the spherical harmonic functions used in this work
\begin{eqnarray}
Y_{0}^{0}(\theta_K,\phi) =\frac{1}{\sqrt {4\pi}},\;\;\;
Y_{1}^{0}(\theta_K,\phi) =\sqrt{\frac{ 3}{ {4\pi}}} \cos\theta_K,\;\;\;
Y_{1}^{\pm 1}(\theta_K,\phi) =\mp \sqrt{\frac{ 3}{ {8\pi}}} \sin\theta_K,\nonumber\\
Y_{2}^{0}(\theta_K,\phi) =\sqrt{\frac{ 5}{ {16\pi}}} (3\cos^2\theta_K-1),\;\;\;
Y_{2}^{\pm 1}(\theta_K,\phi) =\mp \sqrt{\frac{ 15}{ {32\pi}}} \sin(2\theta_K),\nonumber\\
Y_{3}^{0}(\theta_K,\phi) =\sqrt{\frac{ 7}{ {16\pi}}} (5\cos^3\theta_K-3\cos\theta_K),\;\;\;
Y_{3}^{\pm 1}(\theta_K,\phi) =\mp \sqrt{\frac{ 21}{ {64\pi}}} \sin \theta_K (5\cos^2\theta_K-1),\nonumber\\
Y_{4}^{0}(\theta_K,\phi) ={\frac{ 3}{ {16 \sqrt\pi}}} (35\cos^4\theta_K-30\cos^2\theta_K+3),\;\;\;
Y_{4}^{\pm 1}(\theta_K,\phi) =\mp {\frac{ 3\sqrt 5}{ {8\sqrt\pi}}} \sin \theta_K (7\cos^3\theta_K-3\cos\theta_K).
\end{eqnarray}
\end{itemize}

\end{appendix}



 
\end{document}